\documentclass[%
,twocolumn%
,amssymb, amsmath,nobibnotes, aps, pre,longbibliography]{revtex4-2}
\usepackage{bm}%
\expandafter\ifx\csname package@font\endcsname\relax\else
 \expandafter\expandafter
 \expandafter\usepackage
 \expandafter\expandafter
 \expandafter{\csname package@font\endcsname}%
\fi

\usepackage{graphicx,subfigure}
\usepackage{amssymb,amsmath,wasysym}
\DeclareGraphicsExtensions{.pdf,.png}
\graphicspath{{./figs/}{./old_figs/}}
\usepackage[usenames,dvipsnames]{color}   % if you want to highlight some part of the paper
\newcommand{\tcr}[1]{\textcolor{black}{#1}}
\definecolor{BLACK}{rgb}{0,0,0} % needed for \tcr in section titles (revtex uppercases them)

\usepackage[normalem]{ulem} % to bar text for instance

\begin{document}

\title{Spontaneous polarization for protrusion-driven cell crawling}

\author{P. Recho }
\affiliation{
LIPhy, CNRS--UMR 5588, Universit\'e Grenoble Alpes, F-38000 Grenoble, France
}
\email{pierre.recho@univ-grenoble-alpes.fr}

\date{\today}%

\begin{abstract} \small 
We propose a minimal one-dimensional continuum model for the spontaneous initiation of protrusion-driven cell crawling on a rigid substrate. The cell cytoskeleton is represented as a viscous actin meshwork that turns over in the bulk and polymerizes at two moving cell edges. Symmetry breaking  arises from the feedback between cell motion, an external chemical regulator of actin nucleation, and actin polymerization at the cell fronts. When the cell moves, the regulator becomes polarized around the moving boundaries, thereby imposing different actin nucleation densities at the two edges. This generates unequal protrusive rates, which in turn reinforce motion and sustain the chemical polarization. Above a critical protrusive activity, the static symmetric state loses stability and the system undergoes a bifurcation toward a motile polarized state. Depending on how the external cue controls actin nucleation, the transition can be either supercritical or subcritical, leading in the latter case to coexistence between static and motile states. Using parameter values appropriate for keratocyte cells, the model predicts \tcr{crawling speeds of the correct order of magnitude and actin-density profiles that are qualitatively similar to experimental observations}, including asymmetric edge-localized density peaks. These results identify a generic mechanism by which external biochemical regulation of actin nucleation can trigger spontaneous motility along a one-dimensional track without requiring molecular motors, specific adhesion dynamics, deformable substrates, or pre-existing polarity. 

\end{abstract}

\maketitle

\section{Introduction}

Cell motility is a crucial process that underpins many biological functions and pathologies in organisms. More specifically, the conversion of a cell from a static state into a motile one  constitutes a decisive \tcr{change of behavior} that integrates external signaling, transcriptional reprogramming, and cytoskeletal remodeling. For instance during the epithelial-to-mesenchymal transition, transcription factors cooperate with Rho-family GTPase complexes to polarize the cytoskeletal filaments and create a front--rear polarity \cite{de2018optogenetic}. Such transitions also occur, for instance,  in wound re-epithelialization by keratinocytes \cite{raja2007wound},  when neutrophils transition from a spherical to a polarized morphology after engaging adhesive proteins on an endothelium \cite{roberts2019neutrophil} or when malignant cells start to metastasize.
 
A number of biophysical models treating the cell as a continuum medium have addressed such transition from a static to a motile state and vice versa as a spontaneous symmetry breaking instability of the cell chemo-mechanical machinery \cite{callan2016actin}. The instability typically occurs when a control parameter linked to a cellular active process reaches a critical value. Non-exhaustive examples include the emergence of oriented friction stemming from the microscopic attachment/detachment dynamics \cite{barnhart2015balance,sens2020stick,ron2020one}, the auto-organization of molecular motors in a viscous active gel representing the cell cytoskeleton \cite{recho2013contraction, callan2013active,lavi2020motility,drozdowski2023optogenetic,winkler2025active}, the coupling between cell shape and force production \cite{ziebert2012model,camley2013periodic, ziebert2012model,blanch2013spontaneous,schmidt2025myosin}, a splay instability in the nematic orientation of the cytoskeletal filaments \cite{tjhung2012spontaneous,giomi2014spontaneous} or a wave-pinning instability in the reaction-diffusion dynamics of Rho-family GTPases controlling the cytoskeleton mechanics \cite{mori2008wave,vanderlei2011computational,copos2020hybrid}. 

\tcr{Our aim is to study, in quantitative detail, a mechanism of spontaneous polarization based on a well-established active function of the cell cytoskeleton: the capacity of its actin meshwork to grow against the cell membrane} \cite{PollardEarnshaw2017}. The complementary functions of adhesion to the substrate and contraction of the meshwork by molecular motors will deliberately be ignored. Incorporating this property into cell motility models is not new. Mogilner and Oster \cite{mogilner1996cell} extended the Brownian ratchet model of Peskin et al. \cite{peskin1993cellular}, which describes the protrusive rate of a growing actin filament against a thermally fluctuating membrane, to deformable filaments in order to model lamellipodial protrusion and Listeria propulsion. These protrusive rates were integrated in one-dimensional (1D) models of cell motility \cite{larripa2006transport,juelicher2007active,recho2016maximum,abeyaratne2024continuum}. In general, however, they are imposed asymmetrically between the front and the rear of the cell and there is no spontaneous transition from a static to a motile state. Sometimes the asymmetry is not directly imposed but is inherited from a polarized chemical system that regulates the protrusion rate of the cytoskeleton at the membrane \cite{ambrosi2016mechanics,giverso2018mechanical}.

The question of a spontaneous polarization of the cytoskeleton itself due to actin polymerization/depolymerization has also been addressed. The paradigmatic example of Listeria propulsion \cite{prost2008physics} was used in \cite{van2005stress,john2008nonlinear} to expose a curvature-driven geometric instability whereby actin grows as a shell on a closed curved surface. Because that growth cannot be accommodated without building incompatible elastic stresses in the meshwork, any small geometric asymmetry in shell thickness or shape is amplified until the symmetric shell breaks into a polarized comet tail. Another idea was put forward in \cite{blanch2013spontaneous}, which describes keratocyte-fragment motility as a geometric instability. Actin polymerization at the boundary of an initially circular viscous meshwork couples growth to shape deformations, so that a small asymmetry in the contour reorganizes growth and the internal flow to break the circular symmetry. Yet another example is presented in \cite{schmidt2025myosin} where actin polymerization alone can generate spontaneous cell polarity.  In this model, spontaneous polarization arises again because normal actin growth on a curved membrane generates curvature-dependent tangential stresses. Any small actin filament number density asymmetry creates a stress gradient that drives cortical flow reinforcing that asymmetry. In these three examples, the nonlinear coupling between the actin growth and the curved cell shape plays a crucial role and these paradigms are not applicable to the 1D situation of a cell moving along a track. We recently showed in \cite{etienne2024initiation} that even a 1D protrusion-driven crawling model can undergo spontaneous polarization and motion from a symmetric protrusive activity, but only because the cell is mechanically coupled to a deformable elastic substrate. Then the substrate displacement follows the cell traction forces and feeds back nonlocally on them, so symmetry breaking is essentially mediated by the substrate deformation.

In contrast to these examples, the present model can already explain the static to motile transition in 1D on an undeformable substrate. The contact between the cell and the substrate is also made completely nonspecific. The crucial ingredient at the origin of the spontaneous polarization of the cell is the interplay between an external chemical agent regulating the nucleation density of the actin at the membrane and the internal flow of actin that pushes the cell fronts.  If the cell moves in one direction, it polarizes the regulatory cue which controls the nucleation density of actin. This density feeds back on the protrusive rates at the cell fronts, whose mismatch controls the cell crawling direction and speed. Enough polarization of the cue is necessary to initiate motion, which itself polarizes the cue. The resulting bifurcation between a static and a motile state can be supercritical or subcritical depending on how the external chemical agent controls the front nucleation density. \tcr{Although} the propulsion mechanism is completely different, the instability presented in this paper is conceptually similar to that described for autophoretic particles \cite{michelin2013spontaneous, moran2017phoretic} where the flow created by the particle motion creates a front-back concentration asymmetry which further supports the motion.

Importantly, the aim of this model is not to quantitatively capture a specific experimental situation but rather to expose the minimal ingredients of a generic paradigm which can apply to varying degrees to explain the crawling motion of different cell types. We shall nevertheless show that this paradigm is applicable to the well-documented context of keratocyte cells  since it involves physical coefficients with realistic orders of magnitude.

\section{Model formulation}

\begin{figure}[h!]
\centering{\includegraphics[width=0.5\textwidth]{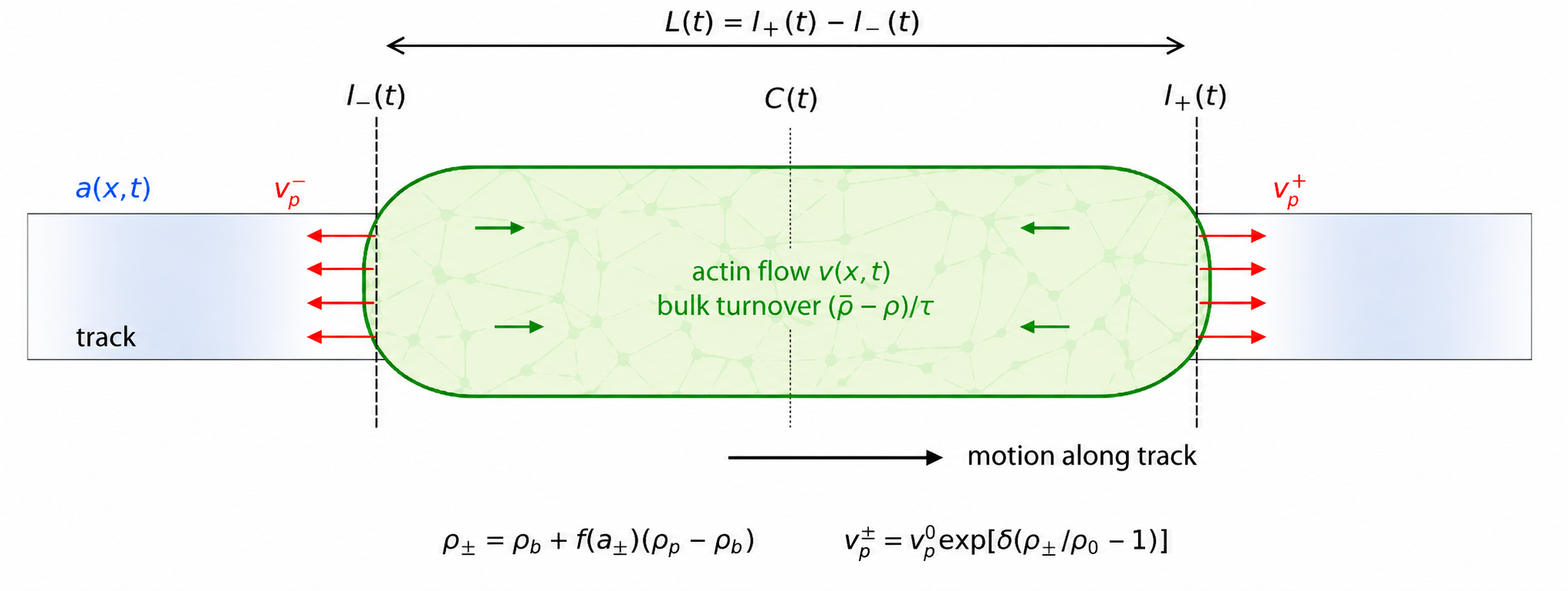}}
\caption{\label{fig:scheme_model} Scheme of the model of an active gel with internal turnover and protrusive rates controlled by an external signaling molecule.}
\end{figure}

To mechanically model the actin cytoskeleton of a moving cell, we consider a thin layer of active gel \cite{kruse2005generic} with material points labeled by the spatial coordinate $x\in [l_-(t),l_+(t)]$ between two moving fronts $l_{\pm}(t)$ where $t\geq 0$ denotes the time. See Fig.~\ref{fig:scheme_model} for a scheme of the model and the introduced notations. We consider one of the simplest constitutive behaviors of the gel 
\begin{equation}\label{e:cst_be}
\sigma=\eta \partial_xv+k\frac{L-L_0}{L_0},
\end{equation}
where $\sigma(x,t)$ is the total axial stress in the layer, $\eta$ is the gel viscosity which accounts for turnover of crosslinkers connecting the biopolymer filaments, $v(x,t)$ is the gel velocity and $k$ is an effective cell elasticity \cite{recho2013asymmetry, putelat2018mechanical} that controls the gel length $L=l_+(t)-l_-(t)$ close to a reference length $L_0$. We purposely neglect the influence of the heterogeneous distribution of active stress in the gel, which is known to create self-polarization by itself \cite{recho2013contraction}. \tcr{Possible origins of an active stress in the absence of molecular motors are discussed in Appendix~A.} 

%We shall discuss the integration of this effect later in the text but the simpler model only requires a protrusive driving.   

Additionally, in order to consider an unspecific interaction with the substrate, we consider the case where the force balance takes the form $\partial_x\sigma=0$ and 
\begin{equation}\label{e:global_fb}
\int_{l_-}^{l_+}v(x,t)\mathrm{d}x=0.
\end{equation} 
This situation appears as a limiting case of the setting where traction forces exerted on the substrate  take the form of an effective friction \cite{tawada1991protein}: $\partial_x\sigma=\xi v$, where $\xi$ is the sliding friction  coefficient. When  the frictional dissipation is small  compared to the viscous dissipation, we obtain the \tcr{aforementioned} contact model, see \cite{chelly2022cell2} for a formal derivation. \tcr{Since $\partial_x(\eta\partial_x v)=\xi v$, this limit is valid when the cell length is small compared with the hydrodynamic length $\sqrt{\eta/\xi}$, i.e. $\xi L_0^2/\eta\ll 1$. The applicability of this approximation is discussed in \cite{chelly2022cell2}. For keratocytes, it requires substrates of low adhesion; see \cite{barnhart2011adhesion} for a quantification of $\xi$. The influence of a finite friction is discussed in Sec.~\ref{sec:ext}.} This setting is generic in the sense that it uncouples the internal active layer mechanics from the nature of the substrate. Since we also do not consider any external traction exerted at the gel moving fronts, the stress $\sigma$ vanishes uniformly in the gel and \eqref{e:cst_be} leads to the following expression of the velocity field:
\begin{equation}\label{e:vel}
v(x,t)=\frac{-k(L(t)-L_0)}{L_0\eta} (x-C(t)),
\end{equation}  
where $C(t)=(l_+(t)+l_-(t))/2$ denotes the center of the active gel layer.  This velocity field can be related to the experimentally measured retrograde flow of actin \cite{vallotton2005tracking,chelly2022cell}.

We now turn to the modelling of the material turnover in the gel. Denoting $\rho(x,t)$ the mass density of the gel, \tcr{the mass balance} reads \cite{kruse2005generic}
\begin{equation}\label{e:mass:bal}
\partial_t\rho+\partial_x(\rho v)=\frac{\bar{\rho}-\rho}{\tau},
\end{equation}
where $\bar{\rho}$ is an equilibrium density corresponding to a balance between polymerization and depolymerization in the bulk and $\tau$ is the typical timescale of the material turnover in the bulk. At the gel fronts, actin filaments are nucleated and grow at a certain rate. Thus, the mass fluxes at the sides
\begin{equation}\label{e:mass_fluxes}
 m_{\pm}=\rho(l_\pm(t),t)(v(l_\pm(t),t)-\dot{l}_{\pm}(t))
\end{equation}
are imposed \cite{mogilner2002regulation}, where $\dot{l}_{\pm}=\partial_tl_{\pm}$ denotes the velocities of the moving fronts. As we impose polymerization at the fronts corresponding to an incoming mass flux at both sides, $m_+\leq 0$ and $m_-\geq 0$. As a result, the characteristics  are inward at both fronts \cite{GodlewskiRaviart1991} and the density therefore needs to be imposed at the membrane:
\begin{equation}\label{e:front_densities}
 \rho_{\pm}(t)=\rho(l_\pm(t),t).
\end{equation} 
\tcr{Biologically}, the values $\rho_{\pm}$ are controlled by the concentration of nucleators (such as Arp2/3, for instance \cite{svitkina1999arp2}).

Defining the polymerization velocities at the fronts as $v_p^{\pm}=\mp m_{\pm}/\rho_{\pm}\geq 0$,
we assume $v_p^{\pm}$ to depend on the gel density in the following way:
\begin{equation}\label{e:vp}
v_p^{\pm}=v_p(\rho_{\pm})=v_p^0\exp(\delta (\rho_{\pm}/\rho_0-1))\geq 0,
\end{equation}
expressing the fact that, from its value $v_p^0$ when $\rho=\rho_0$,  the protrusion velocity can increase with an increasing density if $\delta\geq 0$ or decrease if $\delta\leq 0$. Considering the elastic stress in the biopolymer meshwork $-E(\rho/\rho_0-1)$, where $E$ is the Young modulus of the meshwork and $\rho_0$ can be interpreted as the density at which there is no elastic stress in the meshwork, Kramers rate theory   suggests that $v_p=v_p^0\exp(E(\rho/\rho_0-1)/\sigma_*)$ where $\sigma_*\geq 0$ represents the scale of thermal fluctuations in the kinetic process of monomer attachment to an existing nucleating site \cite{plastino2004effect}. Thus  $\delta=E/\sigma_*$ leading to $\delta\geq 0$. However, other types of dependencies with $\delta\leq 0$ have also been considered phenomenologically \cite{schmidt2025myosin} based on some experimental observations \cite{colin2023recycling}.  An important assumption in this model is that the concentration of monomers is roughly constant throughout the cell because monomers diffuse rapidly in the cytosol. This condition is indeed met in certain situations such as for fish keratocyte fragments \cite{aroush2017actin}. A more detailed derivation of the gel mechanics is given in Appendix\tcr{~A}.

To close the model, we need a rule that imposes the values $\rho_{\pm}$, and fixes the nucleation densities at both sides.  External chemical signals (growth factors, chemokines) do not directly set the amount of Arp2/3 at the membrane, but they control where and when it becomes active. When a cell senses a signal through surface receptors, this activates small GTPases such as Rac1 and Cdc42 locally at the membrane. These GTPases in turn recruit and activate proteins like WAVE and WASP, which stimulate Arp2/3 to nucleate branched actin filaments and drive protrusion. At the same time, Rac1 also promotes the recruitment of Arpin, an inhibitor that dampens Arp2/3 activity. The result is a locally regulated balance between activation and inhibition, so that external signals are translated into a controlled level of effective Arp2/3 activity at the cell edge \cite{PollardEarnshaw2017}. 

To remain generic, we consider the presence of an external \tcr{regulator} adsorbed on the substrate surface  controlling the nucleation of actin at the membrane (see Fig.~\ref{fig:scheme_model}) and  denote $a(x,t)$ its concentration outside the cell ($x\in ]-\infty, l_-(t)[\cup]l_+(t),\infty[$). \tcr{Possible realizations are growth factors or chemokines interacting with the extracellular matrix \cite{lander2002morphogen}.} It satisfies the \tcr{balance} law:
\begin{equation}\label{e:a_concentration}
\partial_ta-D\partial_{xx} a=\frac{\bar{a}-a}{\tau_a}.
\end{equation} 
The diffusion coefficient of $a$ along the substrate surface is denoted $D$ and $a$ degrades over a timescale $\tau_a$. $\bar{a}$ represents a homogeneous source producing the signal $a$ in a completely unbiased way. We equip \eqref{e:a_concentration} with no-flux boundary conditions at $l_{\pm}$,
$$-a_{\pm}\dot{l}_{\pm}-D\partial_xa_{\pm}=0.$$
\tcr{The regulator is therefore excluded from the region of the substrate covered by the cell: the advancing edge sweeps it ahead while the receding edge uncovers substrate that is depleted in regulator and is progressively replenished by the source and by diffusion.} We could more generally consider a certain flux representing the uptake of \tcr{the regulator} by the cell\tcr{, which would lead to a Robin-type boundary condition,} but for sake of simplicity we assume that the rate of this process is \tcr{slow} enough compared to the renewal kinetics \tcr{so that it can be neglected}. \tcr{The effect of a finite uptake is examined in a more complete 2D setting in Appendix~B.}

We then suppose that the value of $a$ at the fronts fixes the nucleation densities:
\begin{equation}\label{e:nucleation}
\rho_{\pm}=\rho_b+f(a_{\pm})(\rho_p-\rho_b).
\end{equation} 
In \eqref{e:nucleation}, $f$ typically represents the rate of a catalyzing reaction which we simply take of Michaelis–Menten type $f(x)=x/a_*/(1+x/a_*)$, and $\rho_{p,b}$ are two distinct nucleation densities.  When $a_+\gg a_* $, we have that $\rho_+=\rho_p$ while when $a_+\ll a_*$, we have that $\rho_+=\rho_b$. If $a$ is a nucleation inhibitor, we therefore impose $\rho_b>\rho_p$, while if $a$ is a nucleation promoter, we rather expect $\rho_p>\rho_b$. 

Setting $y=(x-\tcr{C}(t))/L$, $s=t/(\eta/k)$, $\rho:=\rho/\rho_0$, $a:=a/\bar{a}$ and nondimensionalizing lengths by $L_0$, the nondimensional model reads:
\begin{subequations}\label{e:nd_model}
\begin{align}
&\partial_s(\rho L)-\partial_y(\rho [L(L-1)y+\dot{C}+y\dot{L}])=\epsilon L(\bar{\rho}-\rho)\label{e:nd_rho}\\
&\partial_s(a L)-\partial_y(a (\dot{C}+y\dot{L})+\frac{d}{L}\partial_ya )=\epsilon_aL(1-a)\label{e:nd_a}\\
& \mp j\exp(\delta(\rho\vert_{\mp 1/2} -1))=\dot{l}_{\mp}\mp L(L-1)/2\label{e:nd_fronts}\\
& \frac{d}{L}\partial_y a \vert_{\mp 1/2}+a\vert_{\mp 1/2} \dot{l}_{\mp}=0\label{e:nd_noflux}\\
&\rho \vert_{\mp 1/2}=\rho_b+\frac{r a\vert_{\mp 1/2}}{1+r a\vert_{\mp 1/2}}(\rho_p-\rho_b).\label{e:nd_nucleation}
\end{align}
\end{subequations}
The model \eqref{e:nd_model} contains several nondimensional parameters: $j=\eta v_p^0/( k L_0)$ represents the base protrusive activity at the gel fronts. $\delta$ is the sensitivity of the protrusive activity on the local density, $\rho_{b,p}:=\rho_{b,p}/\rho_0$ are the ``activated''/``inhibited'' actin nucleation densities at the fronts, $\bar{\rho}:=\bar{\rho}/\rho_0$ is the  turnover equilibrium density in the bulk, $d=\eta D/(k L_0^2)$ represents the diffusion of the \tcr{regulator} $a$, $r=\tcr{\bar{a}}/a_*$ is the ratio between the equilibrium concentration of the \tcr{regulator} outside the cell and the threshold concentration upon which it switches the nucleation density of actin, $\epsilon=\eta/(k \tau)$ is the bulk turnover rate of actin and $\epsilon_a=\eta/(k \tau_a)$ is the typical degradation rate of the \tcr{regulator}. Both $L(t)$ and $V(t)$ are scalar parameters that need to be solved for along with the fields $\rho(y,t)$ (for $y \in [-1/2,1/2]$) and $a(y,t)$ (for $y \in ]-\infty,-1/2[\cup]1/2,\infty[ $). The last four equations \tcr{\eqref{e:nd_a}--\eqref{e:nd_nucleation}} are completely uncoupled from the first one \tcr{\eqref{e:nd_rho}} so that the gel density can be reconstructed a posteriori when the velocity and the length of the gel are known. They would be fully coupled if we would have included some local $\rho$-dependent compressibility of the gel in the constitutive behaviour \eqref{e:cst_be}. See Appendix\tcr{~A}. \tcr{As a consequence, the bifurcation threshold, the motile velocity and the phase diagram presented below do not depend on the bulk turnover parameter $\epsilon$ and $\bar{\rho}$. The symmetry breaking takes place in the subsystem formed by the external regulator and the polymerization at the cell edges, while the bulk turnover only shapes the actin density profile inside the cell.}

A number of nondimensional parameters can be estimated based on existing data. See Table~\ref{t:valpar}.

\begin{table}
\scriptsize
\begin{tabular}{lll}
\hline\hline
Name & Symbol & Typical value \\ 
\hline
Gel viscosity & $\eta$ & $10^5$ Pa s \cite{rubinstein2009actin}\\
Cell effective stiffness & $k$  & $10^4$ $\text{Pa}$ \cite{recho2013asymmetry}\\
\tcr{Regulator} diffusion coefficient & $D$ & $10^{-11}$ $\text{m}^{2}\text{s}^{-1}$ \cite{recho2019theory}\\
\tcr{Regulator} lifetime & $\tau_a$ & $10^{4}$ s \cite{rollins1989environment}\\
Cell length & $L_0$ & $2\times 10^{-5}$ m \cite{rubinstein2009actin} \\
actin polymerization velocity & $v_p^0$ & $0.2\times 10^{-6}$ $\text{m} \text{s}^{-1}$ \cite{rubinstein2009actin} \\
actin depolymerization time & $\tau$ & $4$ s \cite{aroush2017actin} \\
stress sensitivity of polymerization & $\delta$ & $3$  \cite{plastino2004effect} \\
\hline
Characteristic length & $L_0$ & $2\times 10^{-5}$ m  \\
Characteristic time & $t_0=\eta/k$ & $10$ s \\
Characteristic velocity & $v_0=L_0/t_0$ & $2\times 10^{-6}$ $\text{m s}^{-1}$  \\
\hline
Polymerization  & $j=\eta v_p^0/(k L_0)$ & $0.1$ \\
Diffusion & $d=\eta D/(k L_0^2)$ & $0.25$ \\
\tcr{Regulator} turnover& $\epsilon_a=\eta/(k \tau_a)$ & $10^{-3}$\\
Actin turnover& $\epsilon=\eta/(k \tau)$ & $2.5$\\
\hline
\tcr{Regulator/threshold ratio} & \tcr{$r=\bar{a}/a_*$} & \tcr{$0.1$ or $1$ (assumed)}\\
\tcr{Nucleation density (low)} & \tcr{$\rho_b$} & \tcr{$1.2$ (assumed)}\\
\tcr{Nucleation density (high)} & \tcr{$\rho_p$} & \tcr{$1.5$ (assumed)}\\
\tcr{Bulk equilibrium density} & \tcr{$\bar{\rho}$} & \tcr{$0.8$ (assumed)}\\
\hline\hline
\end{tabular}
\caption{\small Estimates of material coefficients (case of keratocyte cells) and nondimensional parameter definitions. \tcr{The last four parameters are not constrained by available data; the values given are those used in Figs.~\ref{fig:bif_diag}--\ref{fig:steady_densities}.}\label{t:valpar}}
\end{table}

\section{Bifurcation between static and motile steady states}

We now analyze the possible steady states of \eqref{e:nd_model} corresponding to $\partial_s\equiv 0$, $L$ constant and $V=\dot{C}$ constant. In this case, \tcr{solving \eqref{e:nd_a} with the boundary conditions \eqref{e:nd_noflux},} the \tcr{regulator} concentration at the cell edges takes the form:
\begin{equation}\label{e:morpho_on_sides}
a\vert_{\mp 1/2}=\frac{\sqrt{4 d \epsilon_a+V^2}\mp V}{\sqrt{4 d \epsilon_a+V^2}\pm V}.
\end{equation}
\tcr{The full profile of the regulator outside the cell reads, with $S=\sqrt{4 d \epsilon_a+V^2}$,}
{\color{black}
$$
a(y)=\left\lbrace
\begin{array}{ll}
1+(a\vert_{-1/2}-1)\,e^{\frac{L(S-V)}{2d}(y+\frac12)} & \text{if } y< -\frac12\\[4pt]
1+(a\vert_{1/2}-1)\,e^{-\frac{L(S+V)}{2d}(y-\frac12)} & \text{if } y> \frac12.
\end{array}
\right.
$$
For $V>0$, the regulator thus accumulates in a layer of size $2d/(S+V)$ (in units of $L_0$) ahead of the leading edge and is depleted in a wake of size $2d/(S-V)$ behind the trailing edge. At high concentrations, such as ahead of the leading edge, diffusion should become nonlinear to account for steric interactions between the adsorbed molecules, which would limit this accumulation. Since our theory is linear, the values of $a$ in these high-concentration regions are not quantitatively realistic. This linear description is nevertheless sufficient to capture the main effect, namely the front--rear polarization of the regulator induced by motion. Moreover, when the regulator can diffuse around the cell, as in the 2D geometry studied in Appendix~B, the front--rear contrast is much weaker while the bias created by motion persists.}
If $V=0$, we have $a\vert_{ 1/2}=a\vert_{-1/2}=1$ and no bias exists in polymerization properties at the cell edges. If the velocity becomes finite (we can consider $V>0$, with $V<0$ being a fully symmetric situation) as $4d \epsilon_a\ll 1$, $a_-\simeq d\epsilon_a/V^2$ and $a_+\simeq V^2/(d\epsilon_a)$, so that $a_-\ll a_*$ and $a_+\gg a_*$  and the nucleation densities roughly become $\rho\vert_{-1/2}\simeq \rho_b$ and  $\rho\vert_{1/2}\simeq \rho_p$. The protrusive rates at the cell edges then become asymmetric and we obtain\tcr{, by summing the two equations \eqref{e:nd_fronts},}
$$V=\frac{j}{2}(\exp(\delta \tcr{(\rho_p-1)})-\exp(\delta \tcr{(\rho_b-1)})).$$
For self-consistency we need $V>0$, this means that either $\rho_p\geq \rho_b$ (i.e. $a$ is an activator of Arp2/3) and $\delta\geq 0$ as we expect from the Kramers rate theory or, $\rho_p\leq \rho_b$ (i.e. $a$ is an inhibitor of Arp2/3) and $\delta\leq 0$ which is a more exotic situation but that cannot be excluded for an active system. Increasing the active driving parameter $j$ leads to a proportional increase of the velocity. The gel length can also be retrieved:
$$L=\frac{1+\sqrt{1+4j(\exp(\delta (\rho_p-1))+\exp(\delta (\rho_b-1)))}}{2},$$
and also increases with $j$ as the protrusive rates push the edges apart. 

\begin{figure}[h!]
\centering{\includegraphics[width=0.85\columnwidth]{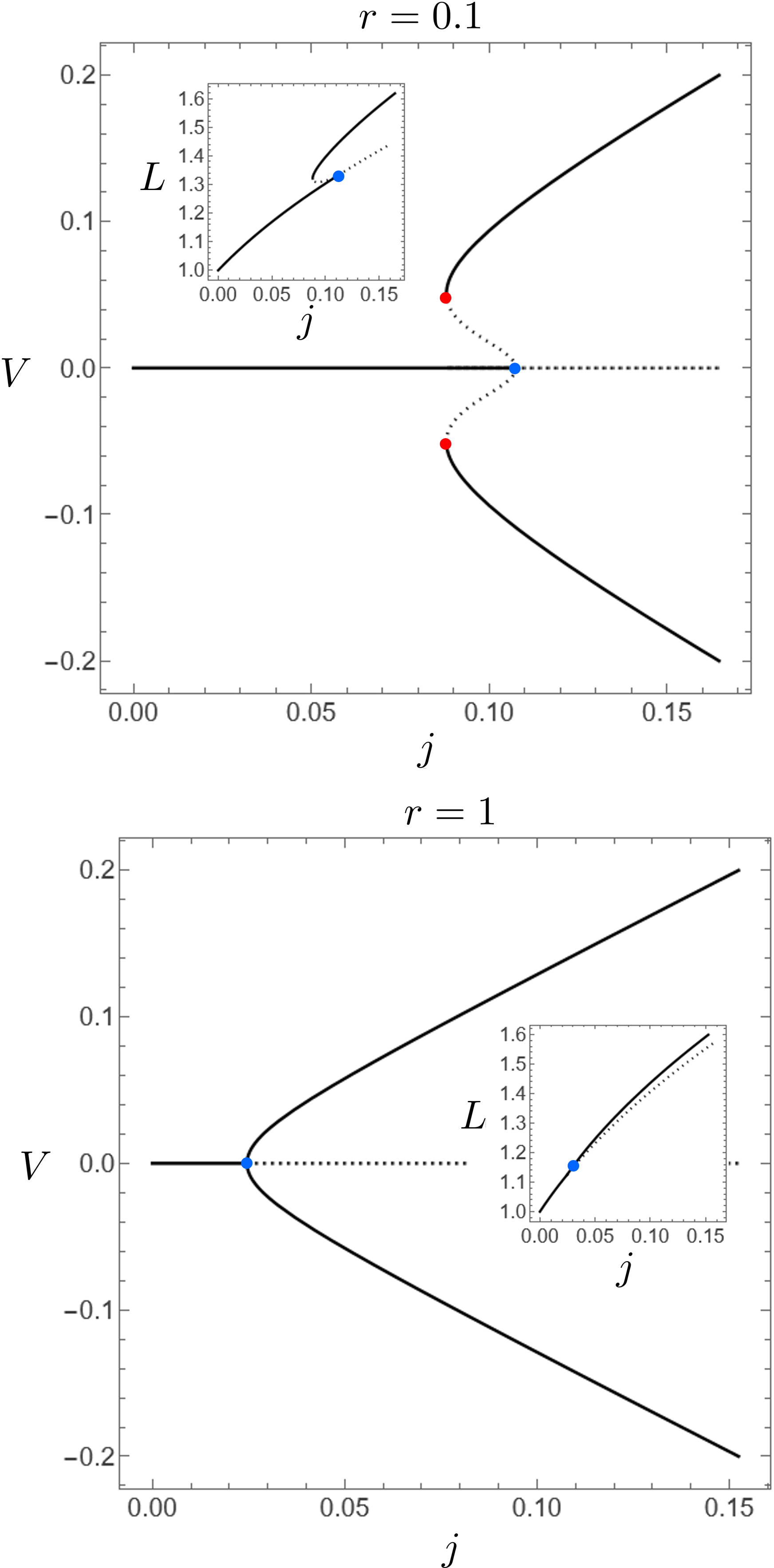}}
\caption{\label{fig:bif_diag} Bifurcation diagrams from the static solution to a motile state upon the increase of the protrusive rate at the cell fronts. The insets show the evolution of the length. Dotted lines indicate unstable states. The bifurcation can be subcritical or supercritical depending on the value of $r$ (see Fig.~\ref{fig:phase_diag}). The blue dots indicate the bifurcation point while the red dots indicate the turning points. \tcr{Here, $j$ is the nondimensional protrusive (polymerization) rate and $r$ is the ratio between the equilibrium concentration of the regulator and the threshold concentration at which it switches the actin nucleation density.} Parameters are $\rho_p = 1.5$, $\rho_b = 1.2$, $d = 0.25$, $\epsilon_a = 0.001$, 
$\delta = 3$. }
\end{figure}

More generally, outside this simple asymptotic case, \eqref{e:morpho_on_sides} \tcr{together with \eqref{e:nd_nucleation}} gives the nucleation densities of actin at the cell edges:
\begin{equation}\label{e:rho_on_sides}
\rho\vert_{\mp 1/2}=\frac{\sqrt{4 d \epsilon _a+V^2} \left(\rho _b+r \rho _p\right)\mp V \left(r \rho _p-\rho _b\right)}{(r+1) \sqrt{4 d \epsilon _a+V^2}\mp (r-1) V}.
\end{equation}
The growth kinetics at the fronts \tcr{\eqref{e:nd_fronts}}
\begin{equation}\label{e:fronts_kinetics}
\mp j\exp(\delta(\rho\vert_{\mp 1/2} -1))=V\mp L(L-1)/2,
\end{equation} 
then provide two algebraic equations that fix the cell velocity $V$ and the gel length $L$.  We show in Fig.~\ref{fig:bif_diag} that two prototypical bifurcations from the static unpolarized state to a motile state can occur at the critical value 
$$j^c=\frac{(r+1)^2 \sqrt{d \epsilon _a} e^{-\frac{\delta  \left(\rho _b+r \left(\rho _p-1\right)-1\right)}{r+1}}}{\delta  r \left(\rho _p-\rho _b\right)}.$$
\tcr{Since the problem is invariant under the transformation $(V,y)\to(-V,-y)$, this bifurcation is a pitchfork in $V$: the static branch $V=0$ exists for all $j$, and the two symmetric motile branches $V>0$ and $V<0$ emerge together at $j=j^c$ and have the same stability.} \tcr{For} the values estimated in Table~\ref{t:valpar} and taking $r=0.1$, $\rho_b=1.2$, $\rho_p=1.5$, we obtain $j^c\simeq 0.107$\tcr{, which is of the same order as the estimated value $j\simeq 0.1$. Since $j^c$ depends strongly on $r$ and on $\rho_p-\rho_b$, which are not constrained by available data, this should only be seen as an order-of-magnitude consistency.}

\begin{figure}[h!]
\centering{\includegraphics[width=0.35\textwidth]{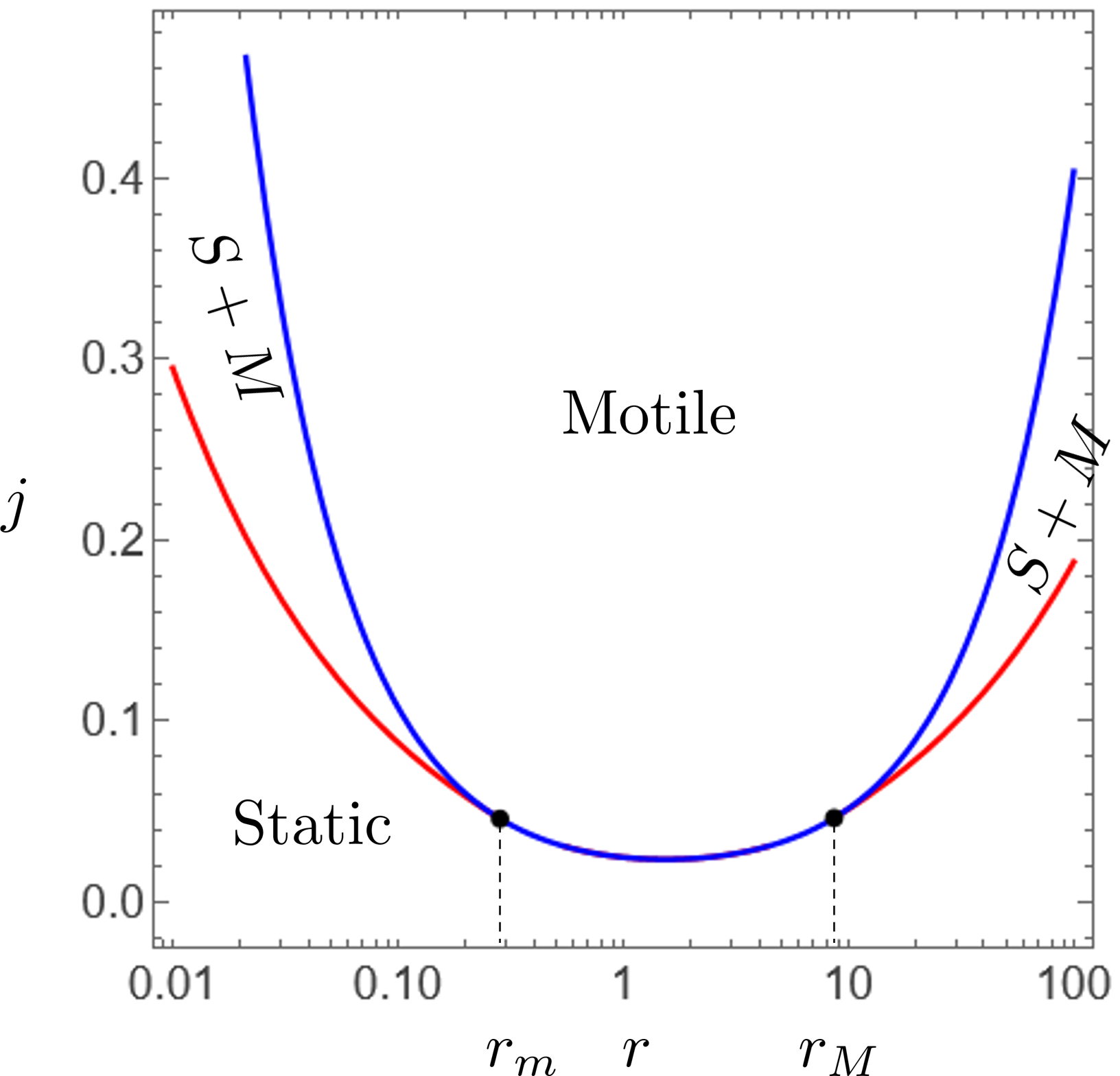}}
\caption{\label{fig:phase_diag} Phase diagram of the system showing in which region of the parameter space $(r,j)$ the system is in a motile state, a static state or where both static and motile states can coexist in a metastable way (M+S). The blue line is the locus of the  bifurcation point shown in Fig.~\ref{fig:bif_diag} and the red line the locus of the turning point when it exists, i.e. outside of the region $[r_m,r_M]$. Interestingly, this region only depends on the positive product of parameters $X=\delta (\rho_p-\rho_b)$. \tcr{Here, $j$ is the nondimensional protrusive (polymerization) rate and $r$ is the ratio between the equilibrium concentration of the regulator and the threshold concentration at which it switches the actin nucleation density.} Parameters are $\rho_p = 1.5$, $\rho_b = 1.2$, $d = 0.25$, $\epsilon_a = 0.001$, $\delta = 3$. }
\end{figure}

If $r\in [r_m,r_M]$ where $r_{m,M}$ are the two positive roots of the equation
$$4 r^2 X^2-12 r \left(r^2-1\right) X+3 ((r-6) r+1) (r+1)^2=0,$$
with $X=\delta (\rho_p-\rho_b)$, then the bifurcation is supercritical while it is subcritical in the other cases. \tcr{For the parameters of Fig.~\ref{fig:phase_diag}, $r_m\simeq 0.28$ and $r_M\simeq 8.6$.} In the supercritical case, the transition from a static to a motile state is smooth while it occurs discontinuously in the subcritical case where there is a region of coexistence of the motile and static states. \tcr{Such a bistability between static and motile states was also obtained in active gel models of contraction-driven motility \cite{chelly2022cell2,drozdowski2023optogenetic}, see also \cite{winkler2025active}.} We show in Fig.~\ref{fig:phase_diag} the related phase diagram in the $(r,j)$ parameter space. In both cases, for $j\simeq 0.1$, the crawling velocity of the gel is of the order of $V\simeq 0.1$ corresponding in dimensional form to $V\simeq 0.2\mu\text{ms}^{-1}$, which is \tcr{of the correct order of magnitude} for our case of keratocyte cells \cite{aroush2017actin}.

Once $V$ and $L$ are determined, the steady state gel density \tcr{given by \eqref{e:nd_rho}} takes the form
\begin{align}\label{eq:rho_ss}
&\rho(y)=\\
&\left\lbrace
\begin{array}{c}
\frac{\bar{\rho}\epsilon}{1-L+\epsilon}+| (L-1) L y+V|^{\frac{\epsilon }{L-1}-1}c_{-} \text{ if } y\leq \tcr{-}\frac{V}{(L-1) L}\\
\frac{\bar{\rho}\epsilon}{1-L+\epsilon}+| (L-1) L y+V|^{\frac{\epsilon }{L-1}-1}c_{+} \text{ if } y\geq \tcr{-}\frac{V}{(L-1) L}.
\end{array}
\right.\nonumber
\end{align}
In \eqref{eq:rho_ss}, $c_{\pm}$ are constants that are set to impose the boundary conditions $\rho \vert_{\mp 1/2}$ imposed in \eqref{e:nd_model}. We also only consider the case where $\epsilon$ is sufficiently large such that $\epsilon>L-1$ and no singularity occurs in the density distribution. Otherwise, a nonlinear model of turnover can be considered to obtain a regular density distribution. See Appendix\tcr{~A}. 

\begin{figure}[h!]
\centering{\includegraphics[width=0.4\textwidth]{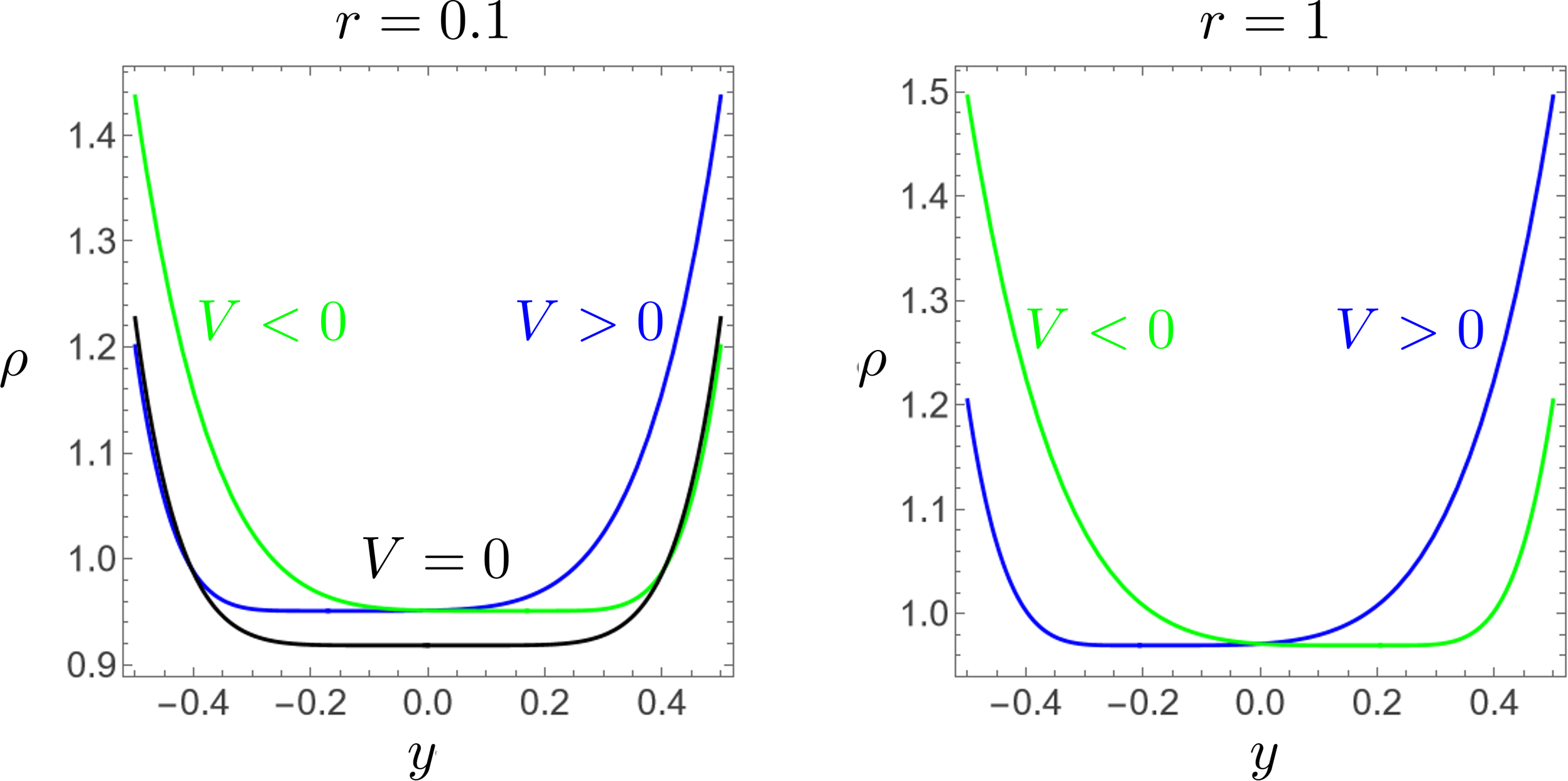}}
\caption{\label{fig:steady_densities} Possible steady state gel densities for $j=0.1$ in the cases $r=0.1$ (subcritical) and $r=1$ (supercritical). In the subcritical case, two symmetric motile states coexist with the static state while in the supercritical case, only the two motile states are stable. Parameters are $\rho_p = 1.5$, $\rho_b = 1.2$, $d = 0.25$, $\epsilon_a = 0.001$, $\delta = 3$, $\epsilon=2.5$ and $\bar{\rho}=0.8$. }
\end{figure}

We show examples of steady state distributions of $\rho$ on Fig.~\ref{fig:steady_densities} corresponding to the cases presented in Fig.~\ref{fig:bif_diag} for the \tcr{estimated} value $j=0.1$.  The static configuration shows two symmetric density peaks at the fronts where polymerization occurs, which then decay in the bulk because of the depolymerization and the internal flow. In the motile state, the distribution becomes biased with the highest density peak being at the leading edge. 
\tcr{These density profiles are qualitatively similar to the experimentally observed ones} \cite{aroush2017actin}. In particular, the presence of two asymmetric peaks at the cell edges in the actin density is captured.

\section{Stability of steady states}

Throughout this text, we have mentioned stability of the steady state, for instance on Fig.~\ref{fig:bif_diag} where the unstable branches are shown in dotted lines while stable branches are shown in full lines. In doing so, we refer to the notion of linear stability of the steady state to small perturbations. That is to say, we inject in \eqref{e:nd_model} the ansatz
\begin{align}\label{e:pertubation_expansion}
a(y,s)=a_{eq}(y)+\theta \delta a (y)\exp(\lambda s)\cos(\omega s)\nonumber\\
\dot{C}(s)=V_{eq}+\theta \delta\dot{C}\exp(\lambda s)\cos(\omega s+\phi_V)\nonumber\\
L(s)=L_{eq}+\theta \delta L\exp(\lambda s)\cos(\omega s+\phi_L),
\end{align}
\tcr{This real form is equivalent to seeking perturbations proportional to $\exp(\Lambda s)$ with a complex eigenvalue $\Lambda=\lambda+i\omega$ and taking their real part. The phases account for the complex amplitudes.}
where $V_{eq}$ and $L_{eq}$ are the solutions of \eqref{e:rho_on_sides}-\eqref{e:fronts_kinetics} shown in Fig.\ref{fig:bif_diag} and $a_{eq}(y)$ is the corresponding steady state solution of the \tcr{regulator} external distribution. We then expand \eqref{e:nd_model} in the small parameter $\theta \ll 1$ to find the perturbations $\delta a (y)$, $\delta\dot{C}$ and $\delta L$. The Fredholm condition ensuring the existence of non-zero perturbation modes gives the growth rate of the perturbation $\lambda$ and the potential oscillations characterized by the angular frequency $\omega$ and the phases $\phi_{V,L}$. The existence of  positive $\lambda$ corresponds to an unstable steady state while if $\lambda$ is negative, the steady state solution is stable.  We do not mention the $\rho$ distribution in the expansion \eqref{e:pertubation_expansion} since it fully uncouples from the rest of the system.  Owing to the simplicity of the (linear) drift term in \eqref{e:nd_model}, all this stability analysis can be carried out analytically\tcr{. As shown in Appendix~C, it reduces to the explicit characteristic equation \eqref{e:C_disp}, and the corresponding computations are provided in the Supplementary Material} \cite{supplementary}. \tcr{The main outcomes are the following. (i) The static state loses stability at $j=j^c$ through a single real eigenvalue associated with a translation of the cell. The bifurcation is therefore a pitchfork without oscillations, and the perturbations that change the cell length remain stable. (ii) For the parameters of Fig.~\ref{fig:bif_diag}, all the unstable eigenvalues are real ($\omega=0$). A motile state is unstable if and only if $j$ decreases with $V$ along its branch, so that stability only changes at the bifurcation and turning points.}

 \section{\tcr{Extensions and related models}}\label{sec:ext}

\tcr{With a finite friction, $\partial_x\sigma=\xi v$, the linear velocity profile \eqref{e:vel} is replaced by
$$
v(x,t)=-\frac{k(L-L_0)}{L_0\eta}\,\ell\,\frac{\sinh\big((x-C)/\ell\big)}{\cosh\big(L/(2\ell)\big)},\qquad \ell=\sqrt{\eta/\xi},
$$
which exhibits exponential boundary layers near the fronts when $L\gg\ell$ and reduces to \eqref{e:vel} when $\ell\gg L$. This velocity field remains odd with respect to the cell center, so that the retrograde flows at the two fronts are opposite. They therefore cancel out when the two equations \eqref{e:nd_fronts} are added, and the cell velocity is still given by $\dot{C}=\frac{j}{2}[\exp(\delta(\rho\vert_{1/2}-1))-\exp(\delta(\rho\vert_{-1/2}-1))]$. Moreover, the drift in \eqref{e:nd_a} only comes from the mapping between the fixed and the co-moving coordinates and does not involve the gel velocity. At steady state, the regulator concentrations at the fronts do not depend on $L$ (see \eqref{e:morpho_on_sides}) and the velocity $V$ is thus independent of $\xi$. As a consequence, the bifurcation diagrams $V(j)$ in Fig.~\ref{fig:bif_diag}, the phase diagram in Fig.~\ref{fig:phase_diag}, and the positions of the bifurcation and turning points are unchanged for $\xi\neq0$. A finite friction only modifies the difference of the two equations \eqref{e:nd_fronts}, i.e. the steady length, which then solves $2(L-1)\,\hat\ell\tanh(L/(2\hat\ell))=j[\exp(\delta(\rho\vert_{1/2}-1))+\exp(\delta(\rho\vert_{-1/2}-1))]$ with $\hat\ell=\ell/L_0$, and, through \eqref{e:nd_rho}, the density profile $\rho$. We chose to work in the limit $\xi\to0$ because the density $\rho$ can then be computed explicitly, which makes the model fully transparent at an analytical level.}

Instead of assuming that $a$ is an external \tcr{regulator},  the effect of $a$ as an internal regulator of the actin nucleation density at the cell edge could be considered. However, if such a species is only diffusing in the cytoplasm and not bound to the actin meshwork, the instability described here does not occur. Keeping the model simple, in this case, for $x\in [l_-(t),l_+(t)]$, we would assume 
\begin{equation*}
\partial_ta+\partial_x(a v^f)-D\partial_{xx} a=0,
\end{equation*}
with no-flux boundary conditions
$$a_{\pm}(v^f_{\pm}-\dot{l}_{\pm})-D\partial_xa_{\pm}=0,$$
where $v^f$ is the velocity of the permeating fluid. Assuming incompressibility of the fluid we have that $\partial_x v^f=0$ and that \eqref{e:cst_be} becomes 
$$
\sigma=\eta \partial_xv+k\frac{L-L_0}{L_0}-p^f,
$$
where $p^f$ is the hydrostatic pressure in the cytosol. If, for simplicity, we do not consider  fluid fluxes through the cell membrane, then $v^f=\dot{l}_{\pm}$ is the common velocity of the cell edges.  Thus, in the coordinate system co-moving with the cell, $a$  follows a diffusive dynamics with fully symmetric boundary conditions. It therefore does not polarize when the cell moves. 

Another simple option, which has been considered in \cite{ron2020one} where $a$ is associated with Arpin is to consider that $a$ is advected with the actin meshwork and follows
\begin{equation}\label{e:drift_diffusion}
\partial_ta+\partial_x(a v)-D\partial_{xx} a=0.
\end{equation}
If we make this change in our model, we recover an instability of the same type as the one described here because a drift remains in \eqref{e:drift_diffusion} even in the co-moving frame since $v$ and $V$ are different.  However, it is unclear in our case why Arpin would be advected with the actin meshwork in normal conditions. In \cite{maiuri2015actin}, the authors make a special optogenetic construction to transiently associate Arpin to actin which otherwise does not directly bind actin. More generally, the instability described in \cite{ron2020one} is different from ours \tcr{because of the origin of the drift of the regulator relative to the cell. There, the regulator is advected by the retrograde flow of actin, and the authors operate in a regime where the cell velocity is much smaller than this flow. In our case, the drift comes from the motion of the fronts themselves. The retrograde flows at the two fronts are opposite and cancel out when the two equations \eqref{e:nd_fronts} are added, so that the cell velocity is set by the difference between the protrusive rates alone. The retrograde flow only sets the cell length. The linear velocity field \eqref{e:vel}, which is odd with respect to the cell center, is a symmetric simplification of the actin flow observed in keratocytes. There, the retrograde flow with respect to the substrate is roughly ten times smaller than the cell velocity  \cite{vallotton2005tracking}.}

It is also interesting to interpret $a$ as molecular motor concentration in  \eqref{e:drift_diffusion} since  motors may mechanically reduce the effective protrusive rate at the cell fronts \cite{bohnet2006weak}. If we additionally include a contractile stress term in \eqref{e:cst_be} that is proportional to $a$:
$$\sigma=\eta \partial_xv+k\frac{L-L_0}{L_0}+\chi a,$$
the resulting model couples a type of protrusion-driven instability similar to the one described here with the contraction-driven instability studied in \cite{recho2013contraction}.
 
\section{Conclusion}

In this work, we have proposed a minimal one-dimensional continuum model for the spontaneous initiation of protrusion-driven cell crawling. The model deliberately removes several mechanisms that can also be at the origin of cell polarization, including molecular-motor contractility, specific adhesion dynamics, substrate deformation, imposed front--rear biochemical polarity, and coupling to cell shape. \tcr{The remaining ingredients are polymerization of a viscous actin meshwork at the two cell fronts and the regulation of actin nucleation by an external chemical. The bulk turnover of the meshwork does not affect the onset of motion but shapes the actin density profile.} This reduction allows to identify a simple feedback loop capable of breaking the left--right symmetry on a rigid 1D track.  Cell motion polarizes the concentration of the external regulator at the two moving boundaries.  This chemical polarization imposes different actin nucleation densities at the two edges. The resulting mismatch in protrusive rates then sustains and amplifies the motion.

This feedback produces a bifurcation from a static symmetric state to a motile polarized state when the protrusive activity exceeds a critical value. Depending on how the external chemical cue controls actin nucleation, the transition can be either supercritical, with a smooth onset of velocity, or subcritical, with coexistence between static and motile states. In the latter case, the model predicts a regime of bistability in which finite perturbations could trigger motility even below the linear instability threshold. For parameter values appropriate to keratocyte cells, the predicted critical protrusive activity \tcr{and crawling speed have the correct order of magnitude, and the actin-density profiles are qualitatively similar to experimental observations}. In particular, the model naturally generates asymmetric actin-density peaks at the cell edges in the motile state, while retaining symmetric edge-localized density peaks in the static state.

The central idea of the model is that chemotaxis need not only be understood as the response of a cell to a preexisting external gradient. Even in an initially homogeneous chemical environment, the fact that a cell boundary moves through a diffusible and degrading species can create a local front-rear imbalance in the concentration sensed by the cell. If this species controls one of the active driving mechanisms that support motion, such as actin nucleation, actin polymerization, contractility, or adhesion turnover, this imbalance can feed back on the velocity itself. This provides a very simple physical mechanism for spontaneous polarization: motion locally reshapes the chemical environment, and the reshaped chemical environment reinforces motion.

Several extensions follow naturally from this picture. First, the external regulator considered here could be replaced by an internal one, provided that it is transported relative to the cell frame, for example by retrograde actin flow.  Second, the chemical feedback could be coupled to other active processes. A regulator that controls molecular-motor contractility would combine the present protrusion-driven instability with the contraction-driven instability. Third, the present one-dimensional framework could be generalized to higher-dimensional geometries, where the same principle may coexist with a cell shape-driven instability. \tcr{As a first step, we show in Appendix~B that the motion of a disc-shaped cell in 2D also polarizes the regulator, with a weaker front--rear contrast than in 1D.}

\section{Appendix\tcr{ A: Thermodynamic derivation}}
 
The aim of this section is \tcr{to} justify our phenomenological model from a thermodynamic perspective and also suggest some routes for its extension/regularisation. We essentially follow our approach developed in \cite{recho2024optimal} but also  consider boundary terms representing protrusion.

The geometric setting is the one already presented in the paper (see Fig.\ref{fig:scheme_model}). The mass \tcr{balance} reads,
\begin{equation}\label{e:m_bal_a}
\partial_t\rho+\partial_x(\rho v)=\tcr{\Gamma}
\end{equation}
with boundary conditions
\begin{equation}\label{e:m_bal_bc_a}
m_{\pm}=\rho_{\pm}(v_{\pm}-\dot{l}_{\pm}),
\end{equation}
where the bulk turnover rate is $\tcr{\Gamma}(x,t)$ and the mass fluxes at the boundaries are $m_{\pm}$.
Momentum balance along the track reads
\begin{equation}\label{e:f_bal_a}
\partial_x\sigma=\xi v,
\end{equation}
with boundary conditions
\begin{equation}\label{e:f_bal_bc_a}
\sigma_{\pm}=\sigma_b=-k\tcr{(L-L_0)/L_0},
\end{equation}
where $\xi$ is the sliding friction coefficient with the solid substrate and $\sigma_b$ the residual stress on the gel boundaries controlling its length.

With these \tcr{balance} laws, we compute:
\begin{itemize}
\item The power of external forces on the gel:
\begin{align}\label{e:work_a}
\frac{\delta W}{\delta t}&=-\int_{l_-}^{l_+}\xi v^2 \mathrm{d}x+\sigma_b(\dot{l}_+-\dot{l}_-)\\
&=\int_{l_-}^{l_+}\sigma\partial_xv\mathrm{d}x-\sigma_b(m_+/\rho_+-m_-/\rho_-),\nonumber
\end{align}

\item The power due to chemical reaction \tcr{controlling} the turnover is
\begin{equation}\label{e:chem_power_a}
\frac{\delta P}{\delta t}=\int_{l_-}^{l_+} \mu_m \tcr{\Gamma} \mathrm{d}x+\mu_b(m_--m_+),
\end{equation}
where $\mu_m$ is the chemical potential controlling the monomer availability in the bulk and $\mu_b$ is the same quantity at the gel boundaries where the membrane is present. 
\item The free energy $F$ of the gel is supposed to depend on $\rho$ only so that
$$F=\int_{l_-}^{l_+}\rho \tcr{\psi}(\rho)\mathrm{d}x,$$
where $\tcr{\psi}$ is the free energy per unit mass. Then
\begin{equation}\label{e:free_ener_var_a}
\frac{ dF}{d t}=\int_{l_-}^{l_+} (\tcr{\Gamma} \mu-p\partial_xv) \mathrm{d}x-m_+\tcr{\psi}_++m_-\tcr{\psi}_-,
\end{equation}
where $\mu=\tcr{\psi}+\rho\partial_{\rho} \tcr{\psi}$ is the Gibbs chemical potential of the gel and $p=\rho^2\partial_{\rho} \tcr{\psi}$ is the pressure in the gel.
\end{itemize}
Next, we combine these quantities to form the dissipation which has to be non-negative by virtue of the second principle,
\begin{align}\label{e:dissip_a}
\mathcal{D}&=\frac{\delta W}{\delta t}+\frac{\delta P}{\delta t}-\frac{dF}{d t}\\
&=\int_{l_-}^{l_+}[(\sigma+p)\partial_xv+(\mu_m-\mu)\tcr{\Gamma}]\mathrm{d}x\nonumber\\
&-(\mu_b-\bar{\mu}_{+})m_++(\mu_b-\bar{\mu}_{-})m_-\geq 0\nonumber,
\end{align}
where $\bar{\mu}_{\pm}=\tcr{\psi}_{\pm}-\sigma_b/\rho_{\pm}$ are the chemical potentials at the protruding boundaries.

Close-to-equilibrium \cite{de2013non}, we write the constitutive relations characterizing the gel mechanical behaviour law and its turnover in the following way:
\begin{align}\label{e:Onsager_a}
&\sigma+p=\eta \partial_xv\tcr{+\zeta(\mu_m-\mu)}\\
&\tcr{\Gamma}=\tcr{\zeta'\partial_xv+}\lambda_m(\mu_m-\mu)\nonumber\\
&\mp m_{\pm}=\lambda_b(\mu_b-\bar{\mu}_{\pm}),\nonumber
\end{align}
where $\eta$ is the gel viscosity, $\lambda_m$ represents a rate of turnover in the bulk\tcr{,} $\lambda_b$ is a rate of turnover at the membrane\tcr{, and $\zeta$ and $\zeta'$ are cross-coupling coefficients between the mechanical and chemical processes in the bulk. These cross-couplings are allowed by symmetry since, in 1D, $\partial_x v$ and $\mu_m-\mu$ are both scalars. The term $\zeta(\mu_m-\mu)$ is an active stress driven by the chemical imbalance of the turnover. Since $\mu$ depends on $\rho$, it would couple the actin density to the gel mechanics. To keep the model as simple as possible, we neglect these cross-couplings in what follows, $\zeta=\zeta'=0$.}

To make these relations explicit, we can for instance assume that
$$\tcr{\psi}(\rho)=\tcr{\psi}_0\left(\frac{\rho_0}{\rho}-1+\log \frac{\rho}{\rho_0}\right),$$
which penalizes deviation of the density from the reference density $\rho_0$ at which pressure vanishes.
We then obtain the pressure $p=\tcr{\psi}_0(\rho-\rho_0)$, so that the Young modulus of the gel can be identified as $E=\rho_0 \tcr{\psi}_0$.  The  bulk turnover rate is $\tcr{\Gamma}=-\lambda_m \tcr{\psi}_0 \log(\rho/\bar{\rho})$ where $\bar{\rho}=\rho_0\exp(\mu_m/\tcr{\psi}_0)$ and finally, $\mp m_{\pm}=\lambda_b(\mu_b-\tcr{\psi}_0(\rho_0/\rho_{\pm}-1+\log \rho_{\pm}/\rho_0)+\sigma_b/\rho_{\pm})$.

\tcr{We obtain the model presented in the paper under the following simplifying assumptions, whose purpose is to keep only the ingredients needed for the mechanism studied here.}
\begin{itemize}
\item \tcr{$E\ll k$: the gel pressure $p=E(\rho/\rho_0-1)$ is negligible compared with the elastic stress $\sigma_b$. Combined with $\partial_x\sigma=0$ (limit $\xi\to 0$), the behaviour law reduces to \eqref{e:cst_be}, where the pressure does not contribute. This is why the density $\rho$ decouples from the mechanics. If this assumption is relaxed, the pressure contributes to the stress. Since the turnover drives the density towards $\bar\rho$, which in general differs from the density $\rho_0$ at which the pressure vanishes, the meshwork is then chemically maintained in a prestressed state. The pressure thus acts as an active stress and couples the density to the mechanics. We studied this situation in \cite{recho2024optimal}.}
\item \tcr{$k\ll \mu_b\rho_0$: the terms $\sigma_b/\rho_{\pm}$ and in $\psi_0$ (since $\psi_0=E/\rho_0$ and $E\ll k$) are negligible compared with $\mu_b$ in the boundary fluxes, so that $\mp m_{\pm}=\lambda_b \mu_b$ is set by the chemical drive at the membrane.}
\item \tcr{The rates of turnover $\lambda_m$ and $\lambda_b$ are $\rho$ dependent, so that the kinetics of renewal depends on the deformation of the meshwork. A Kramers-type dependence for $\lambda_b$ leads to \eqref{e:vp}, and we set $\lambda_m =\rho/( \psi_0 \tau)$ so that \eqref{e:m_bal_a} becomes:}
\end{itemize}
\begin{equation}\label{e:m_bal_a_2}
\partial_t\rho+\partial_x(\rho v)=-\frac{\rho}{ \tau}\log(\frac{\rho}{\bar{\rho}}).
\end{equation} 
Equation \eqref{e:m_bal_a_2} regularizes our classical linear turnover assumption \eqref{e:mass:bal}. Indeed, at steady state, setting $u=\log(\rho/\bar{\rho})$  leads to
$$-\frac{L(L-1)y+V}{L}\partial_yu+\epsilon u=L-1,$$
which modifies \eqref{eq:rho_ss} in the following way:
\begin{align*}
&\rho(y)=\\
&\left\lbrace
\begin{array}{c}
\bar{\rho}\exp(\frac{L-1}{\epsilon})\exp(| (L-1) L y+V|^{\frac{\epsilon }{L-1}}c_{-})\text{ if } y\leq \tcr{-}\frac{V}{(L-1) L}\\
\bar{\rho}\exp(\frac{L-1}{\epsilon})\exp(| (L-1) L y+V|^{\frac{\epsilon }{L-1}}c_{+})\text{ if } y\geq \tcr{-}\frac{V}{(L-1) L}.
\end{array}
\right.
\end{align*}
As $L>1$ because the fronts are pushed by the protrusions, this expression is necessarily regular.

\section{\tcr{Appendix B: Regulator field around a cell moving in 2D}}
\begin{figure}[h!]
\centering{\includegraphics[width=0.48\textwidth]{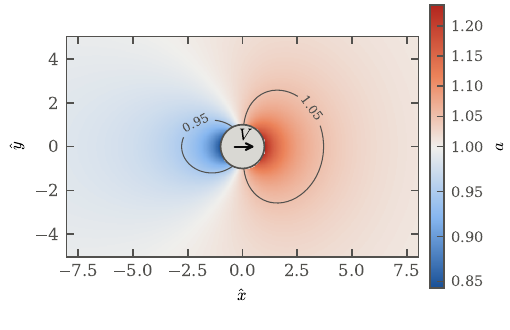}}
\caption{\label{fig:2d_map}\color{black} Steady regulator concentration $a$ around a disc-shaped cell moving to the right, for $\mathrm{Pe}=0.2$, $\mathrm{Da}=10^{-3}$ and $\mathrm{Bi}=0$. The contours show $a=0.95$ and $a=1.05$.}
\end{figure}

\begin{figure*}[t]
\centering{\includegraphics[width=\textwidth]{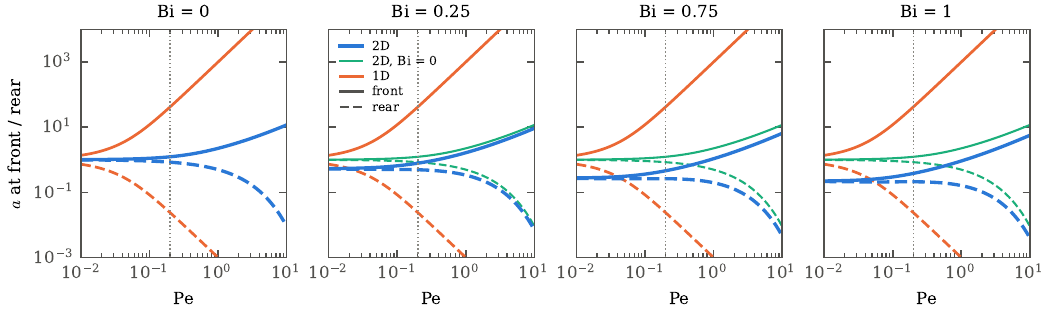}}
\caption{\label{fig:2d_bi}\color{black} Regulator concentration at the front (solid lines) and at the rear (dashed lines) of the cell as functions of Pe, for $\mathrm{Da}=10^{-3}$ and increasing uptake Bi. Blue: 2D disc. Green: 2D disc without uptake. Orange: 1D model, Eq.~\eqref{e:morpho_on_sides}. The dotted vertical line marks the estimated value $\mathrm{Pe}=0.2$.}
\end{figure*}

\begingroup\color{black}
In the main text, the regulator lives on a 1D track and cannot be exchanged with the cell. To assess the robustness of the mechanism with respect to these two simplifications, we consider here a disc-shaped cell of radius $R$ translating at constant velocity $V\mathbf{e}_x$ on a 2D substrate, and we allow for an uptake of the regulator by the cell. In the frame moving with the cell, with origin at its center, the regulator concentration $a(\mathbf{x},t)$ satisfies, outside the disc,
\begin{equation}\label{e:2d_a}
\partial_t a - V\partial_x a - D\Delta a = \frac{\bar a - a}{\tau_a},
\end{equation}
where $\Delta$ is the 2D Laplacian, with $a\to\bar a$ far from the cell and, on the cell edge $|\mathbf{x}|=R$,
\begin{equation}\label{e:2d_bc}
-V a\,\mathbf{n}\cdot\mathbf{e}_x - D\nabla a\cdot\mathbf{n} = -L_p\, a,
\end{equation}
where $\mathbf{n}$ is the outward unit normal to the disc. This Robin condition states that the regulator flux entering the disc equals $L_pa$, where $L_p\geq0$ is the rate of uptake by the cell. For $L_p=0$ it reduces to the 2D counterpart of the no-flux condition used in the main text.

We scale lengths by $R$ and $a$ by $\bar a$. We denote by $(\hat x,\hat y)=\mathbf{x}/R$ the nondimensional Cartesian coordinates and by $(\hat r,\vartheta)$ the associated polar coordinates: $\hat r=|\mathbf{x}|/R$ is the distance to the cell center and $\vartheta$ is the angle to the direction of motion, so that $\vartheta=0$ at the front and $\vartheta=\pi$ at the rear of the cell. Three nondimensional numbers appear:
$$
\mathrm{Pe}=\frac{VR}{D},\qquad \mathrm{Da}=\frac{R^2}{D\tau_a},\qquad \mathrm{Bi}=\frac{L_pR}{D}.
$$
With $R=L_0/2$, they are related to the parameters of the main text by $\mathrm{Pe}=V/(2d)$ and $\mathrm{Da}=\epsilon_a/(4d)$, where $V$ is the nondimensional velocity. The steady problem then reads
\begin{align}\label{e:2d_nd}
&\hat\Delta a + \mathrm{Pe}\,\partial_{\hat x} a + \mathrm{Da}\,(1-a)=0 \quad (\hat r>1),\nonumber\\
&\partial_{\hat r} a + (\mathrm{Pe}\cos\vartheta-\mathrm{Bi})\,a = 0 \quad (\hat r=1),\\
&a\to1 \quad (\hat r\to\infty).\nonumber
\end{align}
For the values of Table~\ref{t:valpar} and the crawling velocity $V=0.1$ (i.e.\ $0.2\,\mu$m\,s$^{-1}$), $\mathrm{Pe}=0.2$ and $\mathrm{Da}=10^{-3}$. The uptake rate is not known, and we vary $\mathrm{Bi}$ between $0$ and $1$.

The solution of \eqref{e:2d_nd} involves the modified Bessel functions of the first and second kind, $I_n(z)$ and $K_n(z)$, of integer order $n\geq0$. The change of unknown $a=1+e^{-\mathrm{Pe}\,\hat r\cos\vartheta/2}\,\Phi$ transforms \eqref{e:2d_nd} into the modified Helmholtz equation $\hat\Delta\Phi-\kappa^2\Phi=0$, with $\kappa=\sqrt{\mathrm{Pe}^2/4+\mathrm{Da}}$. Its solution that is symmetric with respect to the $\hat x$-axis and decay at infinity leads to
\begin{equation}\label{e:2d_sol}
a(\hat r,\vartheta)=1+e^{-\frac{\mathrm{Pe}}{2}\hat r\cos\vartheta}\sum_{n\geq0}A_n\,\frac{K_n(\kappa\hat r)}{K_n(\kappa)}\cos n\vartheta .
\end{equation}
Along the axis, it decays over a length $1/(\kappa+\mathrm{Pe}/2)$ ahead of the cell and over a longer length $1/(\kappa-\mathrm{Pe}/2)$ in its wake.

To determine the coefficients $A_n$, we insert \eqref{e:2d_sol} into the boundary condition. Using the expansion $e^{z\cos\vartheta}=\sum_{m\geq0}\nu_m I_m(z)\cos m\vartheta$, where $\nu_0=1$ and $\nu_{m\geq1}=2$, and $\cos\vartheta\cos n\vartheta=\frac12[\cos(n-1)\vartheta+\cos(n+1)\vartheta]$, and projecting on $\cos m\vartheta$, we obtain the tridiagonal system
\begin{align}\label{e:2d_sys}
&(g_m-\mathrm{Bi})\,A_m+\frac{\mathrm{Pe}}{2\nu_{m-1}}\,A_{m-1}+\frac{\mathrm{Pe}}{4}\,A_{m+1}\nonumber\\
&=\nu_m\Big[\mathrm{Bi}\,I_m\big(\tfrac{\mathrm{Pe}}{2}\big)-\mathrm{Pe}\,I_m'\big(\tfrac{\mathrm{Pe}}{2}\big)\Big],\qquad m\geq0,
\end{align}
with $A_{-1}=0$ and $g_m=\kappa K_m'(\kappa)/K_m(\kappa)<0$. The regulator concentrations at the front and at the rear of the cell are then
\begin{align*}
&a(1,0)=1+e^{-\mathrm{Pe}/2}\sum_{n\geq0}A_n,\\
&a(1,\pi)=1+e^{\mathrm{Pe}/2}\sum_{n\geq0}(-1)^nA_n .
\end{align*}
In practice, we truncate \eqref{e:2d_sys} after $N$ modes. The series converges very rapidly: $N=20$ is sufficient for $\mathrm{Pe}\leq1$ and $N=60$ up to $\mathrm{Pe}=10$. 

Fig.~\ref{fig:2d_map} shows the regulator field for $\mathrm{Pe}=0.2$, $\mathrm{Da}=10^{-3}$ and no uptake. As in 1D, motion creates a front--rear bias: the regulator accumulates ahead of the cell, $a(1,0)=1.23$, and is depleted behind it, $a(1,\pi)=0.84$. This bias is, however, much weaker than in 1D, where the same parameters give $a\vert_{1/2}\simeq42$ and $a\vert_{-1/2}\simeq0.024$ from \eqref{e:morpho_on_sides}. Together with the saturation effects discussed in the main text, this regularizes the large accumulation predicted by the 1D model.

Fig.~\ref{fig:2d_bi} shows the influence of the uptake on the front and rear concentrations over a range of velocities. Uptake lowers the regulator concentration at both ends of the cell, but it preserves the front--rear bias. For $\mathrm{Pe}=0.2$, the ratio $a(1,0)/a(1,\pi)$ even increases from $1.47$ for $\mathrm{Bi}=0$ to $1.81$ for $\mathrm{Bi}=1$.

In all cases, motion thus creates a front--rear bias in the regulator distribution, and this bias is the key ingredient of the feedback loop described in the main text. A quantitative study of the bifurcation in 2D would require coupling this field to a 2D description of the cell mechanics and is left to future work.
\endgroup

\section{\tcr{Appendix C: Linear stability of the steady states}}
\begingroup\color{black}
We streamline here the linear stability analysis summarized in the section on stability of steady states, and we reduce it to an explicit characteristic equation.

The regulator support is two disconnected domains, each bounded by one of the fronts. It is therefore convenient to describe it with the coordinate $z=x-l_+(s)>0$ ahead of the cell and $z=x-l_-(s)<0$ behind it, with lengths scaled by $L_0$. Equations \eqref{e:nd_a} and \eqref{e:nd_noflux} then read
\begin{align}\label{e:C_a}
&\partial_s a-\dot l_\pm\,\partial_z a=d\,\partial_{zz}a+\epsilon_a(1-a),\nonumber\\
&d\,\partial_z a+\dot l_\pm\,a=0\ \ \text{at } z=0,\qquad a\to1\ \ \text{for } |z|\to\infty.
\end{align}
Each domain thus interacts with the rest of the system only through the velocity of its own front. The front velocities follow from \eqref{e:nd_fronts}:
\begin{align}\label{e:C_fronts}
&\dot l_+=j\,e^{\delta(\rho_+-1)}-\frac{L(L-1)}{2},\nonumber\\
&\dot l_-=-j\,e^{\delta(\rho_--1)}+\frac{L(L-1)}{2},\qquad \dot L=\dot l_+-\dot l_-,
\end{align}
where $\rho_\pm=\rho\vert_{\pm1/2}$ depend on the regulator at the fronts through \eqref{e:nd_nucleation}. The density $\rho$ does not enter \eqref{e:C_a}--\eqref{e:C_fronts}. In a steady state, $\dot l_\pm=V$, and the regulator profiles are $a^{eq}_+(z)=1+(a_+-1)e^{-(S+V)z/(2d)}$ and $a^{eq}_-(z)=1+(a_--1)e^{(S-V)z/(2d)}$, with $S=\sqrt{V^2+4d\epsilon_a}$ and $a_\pm=a\vert_{\pm1/2}$ given by \eqref{e:morpho_on_sides}.

 We perturb a steady state as $\dot l_\pm=V+u_\pm e^{\Lambda s}$, $L(s)=L+L_1e^{\Lambda s}$ and $a=a^{eq}_\pm(z)+a_1(z)e^{\Lambda s}$, with $\Lambda\in\mathbb{C}$. Physical perturbations are the real parts of these expressions. Writing $\Lambda=\lambda+i\omega$ gives the form $e^{\lambda s}\cos(\omega s+\phi)$ of \eqref{e:pertubation_expansion}, with growth rate $\lambda$ and angular frequency $\omega$. Ahead of the cell, $a_1$ decays at infinity and satisfies
\begin{align}\label{e:C_lin}
&d\,a_1''+V a_1'-(\Lambda+\epsilon_a)\,a_1=-u_+\,{a^{eq}_+}'(z),\nonumber\\
&d\,a_1'(0)+V a_1(0)+a_+u_+=0 .
\end{align}
Its solution gives the response of the regulator at the front to a perturbation of the front velocity, $a_1(0)=G(\Lambda,V)\,u_+$, with
\begin{equation}\label{e:C_G}
G(\Lambda,V)=\frac{2(S+V)(S+S_\Lambda-2V)}{(S-V)(S+S_\Lambda)(S_\Lambda-V)},
\end{equation}
where $S_\Lambda=\sqrt{V^2+4d(\epsilon_a+\Lambda)}$, with $\mathrm{Re}\,S_\Lambda>0$.  By symmetry, behind the cell $a_1(0)=-G(\Lambda,-V)\,u_-$. Linearizing \eqref{e:C_fronts}, we obtain
\begin{align}\label{e:C_sys}
&(Q_+G_+-1)\,u_+=(L-\tfrac12)\,L_1,\nonumber\\
& (Q_-G_--1)\,u_-=-(L-\tfrac12)\,L_1,\nonumber\\
&\Lambda L_1=u_+-u_-,
\end{align}
where $G_\pm=G(\Lambda,\pm V)$ and $Q_\pm=j\delta\,e^{\delta(\rho_\pm-1)}\,r(\rho_p-\rho_b)/(1+ra_\pm)^2$ measure the sensitivity of the protrusive rates to the regulator at the fronts. Nonzero perturbations exist if and only if $\Lambda$ solves the characteristic equation
\begin{align}\label{e:C_disp}
\Delta(\Lambda)\equiv\ &\Lambda\,(Q_+G_+-1)(Q_-G_--1)\nonumber\\
&-(L-\tfrac12)(Q_+G_++Q_-G_--2)=0 .
\end{align}
A steady state is linearly unstable if and only if \eqref{e:C_disp} has a root with $\mathrm{Re}\,\Lambda>0$. 

\textit{Static state.} For $V=0$, $G(\Lambda,0)=1/\sqrt{d(\epsilon_a+\Lambda)}$ and $Q_+=Q_-=Q_0$, so that \eqref{e:C_disp} factorizes into
$$
\big[Q_0G(\Lambda,0)-1\big]\big[\Lambda\big(Q_0G(\Lambda,0)-1\big)-2L+1\big]=0 .
$$
The first factor corresponds to antisymmetric perturbations, $u_+=u_-$ and $L_1=0$, i.e.\ to a translation of the cell. It has the single root $\Lambda=Q_0^2/d-\epsilon_a$, which is real and becomes positive when $Q_0>\sqrt{d\epsilon_a}$, i.e.\ for $j>j^c$; this recovers the threshold $j^c$ given in the main text. The second factor corresponds to symmetric perturbations, $u_+=-u_-$, which change the cell length. We found that it has no root with $\mathrm{Re}\,\Lambda>0$. The static state therefore loses stability through a real eigenvalue, without oscillations, which is consistent with the pitchfork bifurcation of Fig.~\ref{fig:bif_diag}.

\textit{Motile states.} For the parameters of Fig.~\ref{fig:bif_diag}, all the roots of \eqref{e:C_disp} with $\mathrm{Re}\,\Lambda>0$ are real. Moreover, a motile state is unstable if and only if $j$ decreases with $V$ along the branch, $dj/dV<0$. Stability can thus only change at the bifurcation point and at the turning points. In the supercritical case, the motile branches are stable. In the subcritical case, the part of the branch between the bifurcation point and the turning point is unstable, and the part beyond the turning point is stable. By symmetry, the branches $V>0$ and $V<0$ have the same spectrum.

\bibliography{turnover}

%% \newpage
%% \input{appendixA_force.tex}

%\clearpage

\end{document}